\documentclass[%
 aip,
 amsmath,amssymb,
 reprint,%
]{revtex4-2}

\usepackage{graphicx}
\usepackage{dcolumn}
\usepackage{bm}

\usepackage[utf8]{inputenc}
\usepackage[T1]{fontenc}
\usepackage{mathptmx}
\usepackage{etoolbox}

\usepackage{enumitem}
\usepackage{tikz}

\makeatletter
\def\@email#1#2{%
 \endgroup
 \patchcmd{\titleblock@produce}
  {\frontmatter@RRAPformat}
  {\frontmatter@RRAPformat{\produce@RRAP{*#1\href{mailto:#2}{#2}}}\frontmatter@RRAPformat}
  {}{}
}%
\makeatother

\begin{document}

\preprint{AIP/123-QED}

\title{Classification of Cellular Automata based on the Hamming distance} 

\author{Gaspar Alfaro}
 \email{gaspar.alfaro@urjc.es}
\affiliation{Nonlinear Dynamics, Chaos and Complex Systems Group, Departamento de  Física, Universidad Rey Juan Carlos, Tulipán s/n, Móstoles, 28933, Madrid, Spain}
\author{Miguel A.F. Sanjuán}
 \email{miguel.sanjuan@urjc.es}
\affiliation{Nonlinear Dynamics, Chaos and Complex Systems Group, Departamento de  Física, Universidad Rey Juan Carlos, Tulipán s/n, Móstoles, 28933, Madrid, Spain}

\date{\today}

\begin{abstract}
Elementary cellular automata are the simplest form of cellular automata, studied extensively by Wolfram in the 1980s. He discovered complex behavior in some of these automata and developed a classification for all cellular automata based on their phenomenology. In this paper, we present an algorithm to classify them more effectively by measuring difference patterns using the Hamming distance. Our classification aligns with Wolfram's and further categorizes them into additional subclasses. Finally, we have found a heuristic reasoning providing and explanation about why some rules evolve into fractal patterns.

\end{abstract}


\maketitle

\begin{quotation}

The classification of cellular automata has been a prevalent topic in the field. Wolfram's classification significantly advanced the identification of the main dynamical differences between various automata.  Four different types of dynamical behavior identify the classes of automata whether the system is at equilibrium, is periodic, chaotic or present dissipative and transient complex behavior. Traditionally, determining the class of a rule requires analyzing the space-time dynamics and making a subjective judgment. The difference patterns between two configurations, where an initial cell differs in one configuration, also exhibit distinct dynamical behavior for each class, but the assessment remains subjective. We have discovered that plotting the Hamming distance between the two configurations (i.e., the number of differing cells) over time provides an objective and quicker method for classifying automata into the same four Wolfram classes. If the distance is null, constant or periodic, complex, or shows transient chaos, then the rule belongs to Classes 1 through 4, respectively. Furthermore, we have found a heuristic reasoning explaining the fractal nature of some rules.

\end{quotation}

\section{Introduction}

Cellular automata (CA)~\cite{WolframCA1, WolframCA2} are systems that can be used to produce simple agent-based models exhibiting complex dynamics. Each cell can exist in several possible states and is updated at each time iteration based on the state of neighboring cells, its own state, and one or several update rules. They are employed in various contexts and have broad applications in physics~\cite{PhysicsCA1, PhysicsCA2, PhysicsCA3, PhysicsCA4, PhysicsCA5,PhysicsCA6,PhysicsCA7}, engineering~\cite{EngineeringCA1}, cryptography \cite{CryptographyCA1, CryptographyCA2Lya}, and theoretical biology~\cite{BiologyCA1, BiologyCA2, BiologyCA3}.

Elementary cellular automata (ECA) are the simplest form of one-dimensional CA, with only two possible states for each cell. The state at the next iteration depends solely on the cell itself and its neighbors. Despite their simplicity, some rules exhibit complex behavior and can be used as random number generators.

Wolfram classified Elementary Cellular Automata (ECA) rules into four classes in~\cite{WolframCA_ClassOrigen}, categorizing them based on their temporal behavior, though they are sensitive to the initial conditions. Class-$1$, exhibits uniform behavior, Class-$2$ shows periodic patterns, Class-$3$ displays chaotic dynamics, and Class-$4$ demonstrates transiently complex behaviors.

Wolfram's classification is extensively used today as in~\cite{Koopman}, for example, and other classifications have been developed so far \cite{ClassificationCA1LiPackard1, ClassificationCA2LiPackard2, ClassificationCA3Wuensche, ClassificationCA4TopologicalDynamics, ClassificationCA5, ClassificationCA6, ClassificationCA7}. The authors of \cite{Vispoel} claim that most classification are limited to ECAs. Wolfram's classification, as well as our own, were studied under ECAs but are extensive to other CAs.

When initial conditions are minimally altered by changing one cell, each class responds distinctively, generating unique difference patterns over subsequent iterations. These patterns vary significantly across classes. The Hamming distance, obtained by counting the nonzero elements in these difference patterns, quantifies the degree of change between configurations.

Our study focuses on analyzing the time series of the Hamming distance between successive configurations of ECA. This approach has enabled us to classify ECA rules into a classification scheme that closely aligns with Wolfram's classes and identifies additional subclasses. This breakthrough is significant because it allows us to distinguish between ECA classes based on the time series alone, rather than observing the complex pattern formation behavior, which typically requires more data and is harder to automate.

While we have only analyzed ECA's, we think that the same algorithm can be used for more complex automata, as well as for higher dimensional systems.

This manuscript is organized as follows: In Section~\ref{WolframClassSection}, we introduce the ECAs and present Wolfram's classification of CA. Section~\ref{HammingDistanceSection} outlines our method for classifying ECAs based on the analysis of the Hamming distance in their time series. Finally, we present our main conclusions at the end.

\begin{figure*}
    \centering
    \includegraphics[height=0.75\textheight]{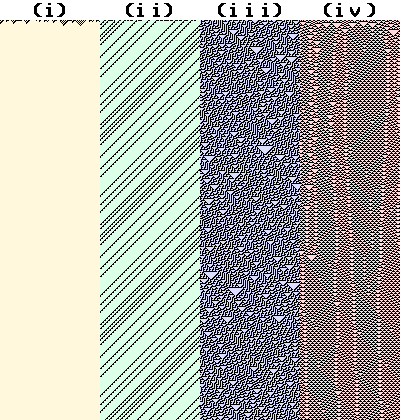}
    \caption{Representation of the four classes of ECA according to Wolfram. The time evolution of the horizontal set of rules is represented downwards as time increases. Different colors represent the state $0$ at each class. (i) Class-$1$ with Rule-$40$, the automaton quickly goes to a fixed state where all cells are $0$. (ii) Class-$2$ with Rule-$6$, the automaton quickly evolves to a periodic state. (iii) Class-$3$ with Rule-$30$, the automaton results in chaotic patterns that does not repeat. (iv) Class-$4$ with Rule-$54$, various distinct regions can be observed before a periodic state is found, for longer times than depicted. The initial cells at the top have a $50\%$ chance of being $0$ or $1$ and the automaton is iterated using a periodic boundary. }
    \label{ClassExamples}
\end{figure*}

\section{Evolutionary Cellular Automata and Wolfram's Classification}
\label{WolframClassSection}

Elementary cellular automata are one-dimensional, binary systems where each cell's state depends only on its own state and that of its immediate neighbors. With each cell having two possible states and considering its three neighbors, there are $2^{2^3} = 256$ possible rules. However, due to symmetry considerations, only $88$ rules are distinct, as listed in Table~\ref{TablaRulesClass} in the Appendix. Each rule exhibits unique behavior from the same initial conditions, though some rules may behave similarly to others.

Wolfram phenomenologically categorized ECA rules into four classes in \cite{WolframCA_ClassOrigen}. These classes correspond to the patterns observed in Fig.~\ref{ClassExamples}, which evolve into uniform, periodic, chaotic, or complex behavior as follows:

\begin{enumerate}[label=(\roman*)]
    \item Class-$1$: Quickly evolves to a fixed state where all states are the same.
    \item Class-$2$: Quickly evolves to a fixed or periodic state.
    \item Class-$3$: Forms a chaotic pattern that does not repeat over time.
    \item Class-$4$: Initially forms complex patterns that, over generally long times, converge into a fixed or periodic state. Because convergence does not occur at the same time everywhere, there are different regions with different behavior, i.e., regions that seem like those at Class-$3$ and others like Class-$2$. The convergence time depends highly on the automaton size.
\end{enumerate}

We present the Wolfram classification, compiled from \textit{WolframAlpha}~\cite{WolframAlpha} in Table~\ref{TablaRulesClass} in the Appendix alongside our own classification. 

\begin{figure}
    \centering
    \includegraphics[width=\linewidth]{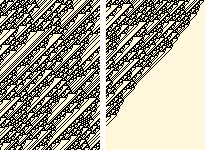}
    \caption{Evolution of a set of cells for Rule-$106$. The horizontal axis represents different cells, and the vertical axis represents time from top to bottom. On the left, evolution under periodic boundary conditions shows patterns that do not repeat for long times, resembling Class-$3$ behavior. On the right, evolution under fixed boundary conditions shows zeros propagating to the left until all cells become zero, aligning with Class-$1$ characteristics.}
    \label{rule106}
\end{figure}

Wolfram's classification applies to other types of CA as well. Certain rules may not exhibit consistent classification when evolved under fixed versus periodic boundary conditions. For instance, Rule-$106$ and its equivalents are categorized as Class-$3$, but behave like Class-$1$ under fixed boundary conditions, as illustrated in Fig.~\ref{rule106}. Under fixed boundaries, zeros propagate until all states become zero. However, under periodic boundaries, the automaton remains chaotic and patterns never repeat except for extremely long timescales, in accordance with Poincaré's recurrence theorem.

Nevertheless, in \textit{WolframAlpha}, Rule-$106$ is classified as Class-$4$, possibly due to the presence of large white spaces. However, we observe that these spaces are diagonal, unlike other Class-$4$ examples that exhibit vertical continuation. Therefore, we argue that the behavior of this automaton aligns more accurately with Class-$3$. Our classification method, presented in the following section, supports this interpretation.

All subsequent calculations are conducted under periodic boundaries with a population size of $100$ cells, unless otherwise specified.

\begin{figure*}
    \centering
    \includegraphics[height=0.75\textheight]{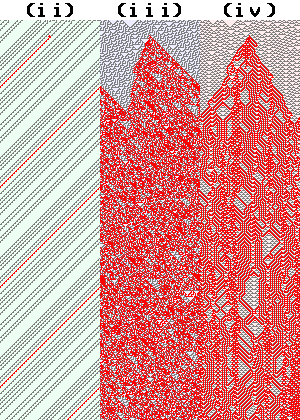}
    \caption{Similar to Fig.~\ref{ClassExamples}, we illustrate the evolution for the different classes with shading. Difference patterns are highlighted in red, demonstrating distinct behaviors for each class.  We have a periodic difference pattern in (ii) for Class-$2$, a noisy behavior in (iii) for Class-$3$ and a complex behavior in (iv) for Class-$4$.
    }
    \vspace{15mm}
    \label{fig:4classDiffPatt}
\end{figure*}

\section{Hamming distance classification}
\label{HammingDistanceSection}

Altering one cell in a cellular automaton can cause the difference to propagate in subsequent iterations. Each class propagates differences in a distinct manner, as shown in Fig.~\ref{fig:4classDiffPatt}. In Class-$2$, the difference pattern is periodic; Class-$3$, it is noisy; and in Class-$4$, it is complex.

The Hamming distance between two strings of equal length is the number of positions where the strings differ. Introduced by Richard Hamming in 1950 in \cite{HammingOrigins}, this metric has been widely used in information and computer science, cryptography, and various other fields. The Hamming distance has been applied in agent-based biological systems to indicate chaotic behavior \cite{HammingChaos1}, \cite{HammingChaos2}. More recently, we extended this approach to social games in \cite{AlfaroHamming}. By analyzing the Hamming distance between two initially identical conditions in the game, one perturbed by an agent with an opposite strategy, we observed that the distance grew in a sigmoid-like curve over time for parameter values indicating spatiotemporal chaos in the \textit{prisoner's dilemma}.

In our case, the Hamming distance between two configurations of cells indicates the number of differing cells. We analyzed the Hamming distance time series of two ECA configurations that initially differ by just one cell. This is equivalent to summing the elements in the difference patterns. The resulting time series behavior varies with the Wolfram class of the rule: it remains constantly $0$ for Class-$1$, evolves to a constant or periodic pattern for Class-$2$, exhibits chaotic or very long periods for Class-$3$, and shows transient chaos for Class-$4$. All ECAs start with random initial conditions, requiring a few iterations before altering the initial cell to ensure accuracy. While some rules exhibit brief transient dynamics critical for classification, approximately $15$ iterations suffice for reliable results.

\begin{figure}
    \centering
    \includegraphics[width=\linewidth]{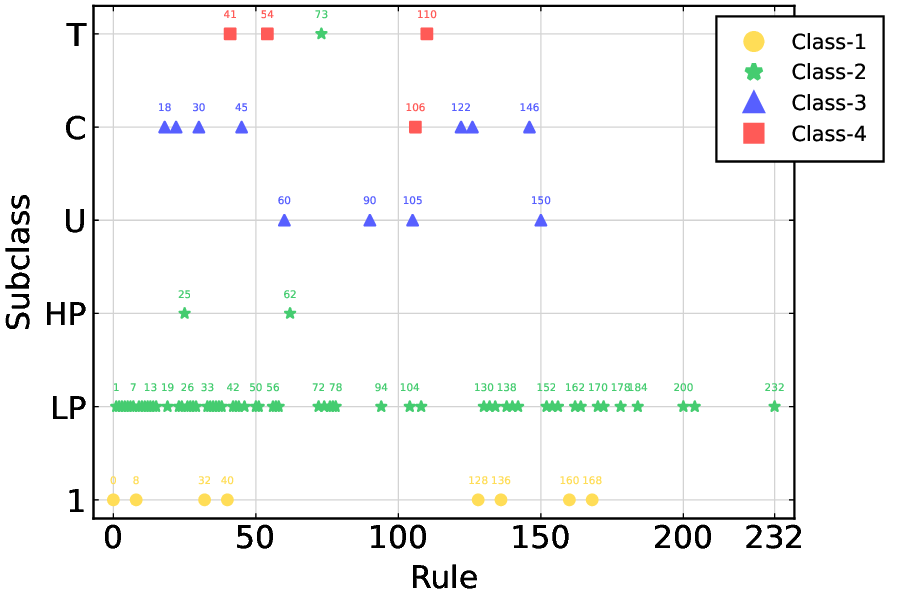}
    \caption{Hamming distance classification. Colors and symbols correspond to Wolfram's classification, while position at the vertical axis correspond to the subclasses.}
    \label{ClassHamming}
\end{figure}

Moreover, we can distinguish subclasses with different behaviors, as seen in Fig.~\ref{ClassHamming}. The time series for a given rule can exhibit different behaviors depending on the initial conditions. We have classified these rules based on the most complex behavior they exhibit across $100$ different initial conditions. The vertical axis of the figure represents a spectrum from simple to complex behaviors. 

We have divided the rules in Class-$2$ into two subclasses. Subclass LP (Low Period) includes rules where the Hamming distance is constant or has a low period $\lesssim 20$ iterations, though most have periods $\le 3$ iterations. Subclass HP (High Period) comprises rules where the Hamming distance has a large period $\gtrsim 5L$ iterations. 

We have also divided the rules in Class-$3$ into two subclasses. Subclass C includes rules where the Hamming distance is chaotic, while Subclass U (Uniform) consists of rules with extremely long periods, where the same pattern repeats across all computations with different initial conditions. Notably, for all initial conditions and rules in Subclass U, the periods are exactly $2046$ iterations, very close to $2^{11} = 2048$. When varying the population size $L$, the periods approximate different powers of $2$. For $L$ values that are a power of $2$, the Hamming distance eventually becomes $0$ after some iterations.

\begin{table}
\begin{tikzpicture}


\fill[blue!14!white] (-0.5*1.5,-0.5) rectangle (3.5*1.5,-1.5);
\fill[blue!14!white] (-0.5*1.5,-2.5) rectangle (3.5*1.5,-4.5);
\fill[blue!14!white] (-0.5*1.5,-5.5) rectangle (3.5*1.5,-6.5);


\draw (0*1.5,0) node {0};
\draw (1*1.5,0) node {0};
\draw (2*1.5,0) node {0};
\draw (3*1.5,0) node {0};

\draw (0*1.5,-1) node {0};
\draw (1*1.5,-1) node {0};
\draw (2*1.5,-1) node {1};
\draw (3*1.5,-1) node {1};

\draw (0*1.5,-2) node {0};
\draw (1*1.5,-2) node {1};
\draw (2*1.5,-2) node {0};
\draw (3*1.5,-2) node {0};

\draw (0*1.5,-3) node {0};
\draw (1*1.5,-3) node {1};
\draw (2*1.5,-3) node {1};
\draw (3*1.5,-3) node {1};

\draw (0*1.5,-4) node {1};
\draw (1*1.5,-4) node {0};
\draw (2*1.5,-4) node {0};
\draw (3*1.5,-4) node {1};

\draw (0*1.5,-5) node {1};
\draw (1*1.5,-5) node {0};
\draw (2*1.5,-5) node {1};
\draw (3*1.5,-5) node {0};

\draw (0*1.5,-6) node {1};
\draw (1*1.5,-6) node {1};
\draw (2*1.5,-6) node {0};
\draw (3*1.5,-6) node {1};

\draw (0*1.5,-7) node {1};
\draw (1*1.5,-7) node {1};
\draw (2*1.5,-7) node {1};
\draw (3*1.5,-7) node {0};

\draw (-0.5,1.35) -- (0.5,1.35);
\draw (-0.5,1.1) -- (0.5,1.1);

\draw (-0.5+1/3,1) -- (-0.5+2/3,1);
\draw (-0.5+1/3,0.75) -- (-0.5+2/3,0.75);

\draw (-0.5,1.35) -- (-0.5,1.1);
\draw (-0.5+1/3,1.35) -- (-0.5+1/3,1.1);
\draw (-0.5+2/3,1.35) -- (-0.5+2/3,1.1);
\draw (-0.5+1,1.35) -- (-0.5+1,1.1);

\draw (-0.5+1/3,1) -- (-0.5+1/3,0.75);
\draw (-0.5+2/3,1) -- (-0.5+2/3,0.75);

 \node (A) at (-0.5,1.35) {};
 \node (B) at (-0.5,1.1) {};
 \node (C) at (-0.5+1/3,1.35) {};

\fill[fill=black] (A.center)--(B.center)--(C.center);

\draw (-0.5+1.5,1.35) -- (0.5+1.5,1.35);
\draw (-0.5+1.5,1.1) -- (0.5+1.5,1.1);

\draw (-0.5+1/3+1.5,1) -- (-0.5+2/3+1.5,1);
\draw (-0.5+1/3+1.5,0.75) -- (-0.5+2/3+1.5,0.75);

\draw (-0.5+1.5,1.35) -- (-0.5+1.5,1.1);
\draw (-0.5+1.5+1/3,1.35) -- (-0.5+1/3+1.5,1.1);
\draw (-0.5+1.5+2/3,1.35) -- (-0.5+2/3+1.5,1.1);
\draw (-0.5+1.5+1,1.35) -- (-0.5+1+1.5,1.1);

\draw (-0.5+1/3+1.5,1) -- (-0.5+1/3+1.5,0.75);
\draw (-0.5+2/3+1.5,1) -- (-0.5+2/3+1.5,0.75);

 \node (A) at (-0.5+1.5+1/3,1.35) {};
 \node (B) at (-0.5+1.5+1/3,1.1) {};
 \node (C) at (-0.5+1.5+2/3,1.35) {};

\fill[fill=black] (A.center)--(B.center)--(C.center);

\draw (-0.5+3,1.35) -- (0.5+3,1.35);
\draw (-0.5+3,1.1) -- (0.5+3,1.1);

\draw (-0.5+1/3+3,1) -- (-0.5+2/3+3,1);
\draw (-0.5+1/3+3,0.75) -- (-0.5+2/3+3,0.75);

\draw (-0.5+3,1.35) -- (-0.5+3,1.1);
\draw (-0.5+3+1/3,1.35) -- (-0.5+1/3+3,1.1);
\draw (-0.5+3+2/3,1.35) -- (-0.5+2/3+3,1.1);
\draw (-0.5+3+1,1.35) -- (-0.5+1+3,1.1);

\draw (-0.5+1/3+3,1) -- (-0.5+1/3+3,0.75);
\draw (-0.5+2/3+3,1) -- (-0.5+2/3+3,0.75);

 \node (A) at (-0.5+3+2/3,1.35) {};
 \node (B) at (-0.5+3+2/3,1.1) {};
 \node (C) at (-0.5+3+1,1.35) {};

\fill[fill=black] (A.center)--(B.center)--(C.center);

\draw (-0.5+4.5,1.35) -- (0.5+4.5,1.35);
\draw (-0.5+4.5,1.1) -- (0.5+4.5,1.1);

\draw (-0.5+1/3+4.5,1) -- (-0.5+2/3+4.5,1);
\draw (-0.5+1/3+4.5,0.75) -- (-0.5+2/3+4.5,0.75);

\draw (-0.5+4.5,1.35) -- (-0.5+4.5,1.1);
\draw (-0.5+4.5+1/3,1.35) -- (-0.5+1/3+4.5,1.1);
\draw (-0.5+4.5+2/3,1.35) -- (-0.5+2/3+4.5,1.1);
\draw (-0.5+4.5+1,1.35) -- (-0.5+1+4.5,1.1);

\draw (-0.5+1/3+4.5,1) -- (-0.5+1/3+4.5,0.75);
\draw (-0.5+2/3+4.5,1) -- (-0.5+2/3+4.5,0.75);

 \node (A) at (-0.5+4.5+1/3,1) {};
 \node (B) at (-0.5+4.5+1/3,0.75) {};
 \node (C) at (-0.5+4.5+2/3,1) {};

\fill[fill=black] (A.center)--(B.center)--(C.center);


\draw (-0.5*1.5,0.5) -- (3.5*1.5,0.5);

\draw (0.5*1.5,1.5) -- (0.5*1.5,-7.5);
\draw (1.5*1.5,1.5) -- (1.5*1.5,-7.5);

\draw (2.45*1.5,1.5) -- (2.45*1.5,-7.5);
\draw (2.55*1.5,1.5) -- (2.55*1.5,-7.5);
\end{tikzpicture}
\caption{Truth table of ECA Rule-$90$. In the first row, the shaded rectangle indicates the referenced cell in each column. Combinations of $0$s and $1$s that produce a $1$ as output are highlighted. }

\label{Rule90}

\end{table}

\begin{figure}
    \centering
    \includegraphics[width=\linewidth]{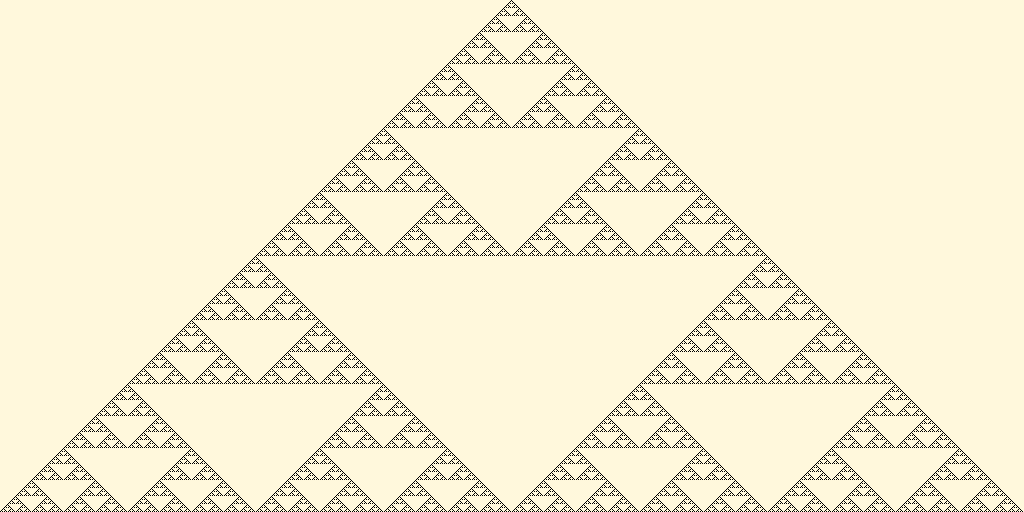}
    \caption{Evolution of the ECA Rule-$90$ of size $1024$ starting with a single black cell, i.e. a $1$, at the top center. The black cell propagates downwards in subsequent iterations, forming the well-known Sierpiński triangle. If a single $0$ is at the beginning, the result is the same except with a black line at the top. This figure exactly matches the difference pattern between two ECA's of the same rule and size, where, after a brief transient (one iteration is enough), the state of the central cell is altered to be a $1$.}
    \label{PatternRule90}
\end{figure}

The fact that the Hamming distance for a rule with chaotic behavior does not depend on the initial conditions is unexpected and contradictory, but it can be explained by analyzing how these rules evolve. For example, Rule-$90$ has an evolution determined by the truth table in Table~\ref{Rule90}. Consider making alterations to the state of the trio of cells. How would these alterations affect the outcome? You could make any alterations from the left side of the table. Choose a row with a $0$ as an output and take the trio of $0$s and $1$s from the left side of the row. We can call that trio A. Adding trio A to any trio B from the left part of the table modulo $2$ results in trio C. The output value for trio C is the same as that for trio B. If the initial trio had a $1$ as its output, the outputs for trios B and C would differ. Thus, the same rule that dictates how a cell evolves also governs how a difference propagates.

This form of self-similarity is encoded in the rule, which occurs exclusively in subclass U. Consequently, when evolving these rules, one would expect to, and indeed does, find fractal patterns.

All rules in the U subclass present fractal patterns in their evolution when the initial conditions consist of a single $1$ in a sea of $0$s, or vice versa. However, some rules outside subclass U, such as Rule-$22$, also exhibit fractal patterns. For these patterns to be true fractals, the size of the automaton must be infinite; otherwise, boundary conditions will limit the scale.

Plotting the difference pattern of Rule-$90$, we obtain Fig.~\ref{PatternRule90}, which exactly matches the evolution pattern of the ECA when starting with a single black cell at the center. This is characteristic of ECAs in subclass U. The familiar Sierpiński triangle pattern, indicative of the chaotic nature of these rules, is evident. Altering which cell is initially permuted will only shift the pattern.

The approximation of the period as powers of $2$ and the reduction in the Hamming distance when $L$ is a power of $2$ can be explained by examining Fig.~\ref{PatternRule90}. The same patterns are scaled and repeated over increasing periods, each multiplied by $2$, due to the self-similarity inherent in the fractal pattern.

Although there are no subclasses for Class-$4$, we denote rules that exhibit transient chaos with a T (Transient), since  there are exceptions.

\begin{figure}
    \centering
    \includegraphics[width=\linewidth]{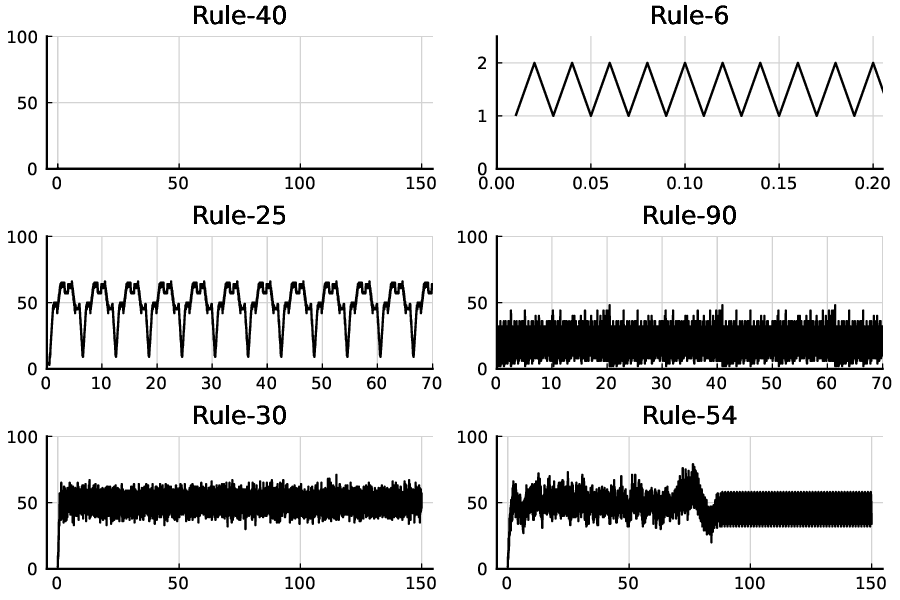}
    \caption{For 6 different rules belonging to: Class-$1$ (Rule-$40$), Class-$2$ subclass LP (Rule-$6$), Class-$2$ subclass HP (Rule-$25$), Class-$3$ subclass U (Rule-$90$), Class-$3$ subclass C (Rule-$30$) and Class-$4$ subclass T (Rule-$54$); we represent the Hamming distance over time. The horizontal axis is in units of $L=100$ iterations, which is the size of the population.}
    \label{HammingDistanceYDifferencePattern6classes}
\end{figure}

In Fig.~~\ref{ClassHamming} and Table~\ref{TablaRulesClass}, we present all the different rules in each subclass. In Fig.~\ref{HammingDistanceYDifferencePattern6classes}, we show the Hamming distance of $6$ different rules corresponding to each subclass. We observe that rules belonging to a Wolfram class, as cataloged in \textit{WolframAlpha}, are grouped into one or two specific subclasses within that class, with two exceptions.

Rule-$106$, which we mentioned in the previous section, is classified in \textit{WolframAlpha} as Class-$4$. Although its pattern of evolution is similar to that of Rule-$3$, the lack of transient behavior and the presence of endless chaos suggest that Rule-$106$ does not properly belong to Class-$4$. It would be more appropriate to classify it as Class-$3$.

\begin{figure}
    \centering
    \includegraphics[height=0.95\textheight]{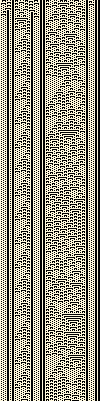}
    \caption{Time evolution of Rule-$73$ over $400$ iterations. The different cells are represented on the horizontal axis and time on the vertical axis from top to bottom. Besides, being classified as Class-$2$ by Wolfram, this pattern belongs to Class-$4$ more accurately.}
    \label{rule73Behaviour}
\end{figure}

The other exception is Rule-$73$. The Hamming distance for this rule shows transient chaos before stabilizing periodically, placing it in subclass T. However, this behavior was observed in only one of the $100$ computations of the Hamming distance. In Fig.~\ref{rule73Behaviour}, we illustrate its pattern of evolution. Periodic and chaotic regions are visible, and iterating further reveals that the chaotic region eventually becomes periodic, thereby classifying it as Class-$4$. 

Additionally, rules classified as subclass HP from Class-$2$ could also be considered for inclusion in subclass T and Class-$4$. However, the instances of complex behavior are very brief, and none of the Hamming distance computations exhibited transient chaos.

Given a rule, the time series obtained by measuring the Hamming distance between two configurations initially differing by just one cell can exhibit varying behavior depending on the initial conditions and which cell is altered initially. For instance, a rule in subclass T may yield a time series resembling subclass LP, but only rules in subclass T exhibit transient chaos. It is emphasized that for the classification used here, focus is placed on time series exhibiting more complex behavior.

\section{Conclussions}

Wolfram classified CA rules by observing the evolution of various initial conditions and discerning patterns from images. In contrast, our developed algorithm efficiently classifies ECA rules by analyzing time series data. Our approach aligns closely with Wolfram's classifications, except for some rules that are arguably misclassified in the \textit{WolframAlpha} engine. Wolfram himself acknowledged that his classification system could vary depending on different initial conditions.

Our classification method relies on measuring the Hamming distance between two initially close configurations, which we analyze based on their periodicity or complexity. We classify rules based on the resulting time series as follows: rules that produce a periodic time series are categorized as Class-$2$, those yielding chaotic or long-period time series as Class-$3$, and rules exhibiting transient chaos as Class-$4$. Trivial rules that result in a Hamming distance value of $0$ are classified as Class-$1$.


\section*{Acknowledgments}

This work has been supported by the Spanish State Research Agency (AEI) and the European Regional Development Fund (ERDF, EU) under Project No.~PID2019-105554GB-I00 (MCIN/AEI/10.13039/501100011033).


\appendix

\section{}

The following table shows all distinct ECA rules, their equivalents, and their classification as compiled from \textit{WolframAlpha} and our subclassification.

\setlength{\tabcolsep}{12pt}
\setlength{\tabcolsep}{12pt}
\begin{table*}
    \caption{ECA rules and classification according to Wolfram, and subclasses according to Hamming distance time series analysis. The rules in the equivalent column are the symmetrical counterpart of the rules so they have the same classification. Only the $88$ distinct rules have their own row.}
    \label{TablaRulesClass}
    \begin{tabular}{c|c|c|c|}
    
    Rule & Equivalent & Class & Subclass \\
    \hline
    0 & 255 & 1 & 1  \\
    1 & 127 & 2 & LP  \\
    2 & 16, 191, 247 & 2 & LP  \\
    3 & 17, 63, 119 & 2 & LP  \\
    4 & 223 & 2 & LP  \\
    5 & 95 & 2 & LP  \\
    6 & 20, 159, 215 & 2 & LP \\
    7 & 21, 31, 87 & 2 & LP \\
    8 & 64, 239, 253 & 1 & 1 \\
    9 & 65, 111, 125 & 2 & LP \\
    10 & 80, 175, 245  & 2 & LP \\
    11 & 47, 81, 117 & 2 & LP \\
    12 & 68, 207, 221 & 2 & LP \\
    13 & 69, 79, 93 & 2 & LP \\
    14 & 84, 143, 213 & 2 & LP \\
    15 & 85 & 2 & LP \\
    18 & 183 & 3 & C \\
    19 & 55 & 2 & LP \\
    22 & 151 & 3 & C \\
    23 & - & 2 & LP \\
    24 & 66, 189, 231 & 2 & LP \\
    25 & 61, 67, 103 & 2 & HP \\
    26 & 82, 167, 181 & 2 & LP \\
    27 & 39, 53, 83 & 2 & LP \\
    28 & 70, 157, 199 & 2 & LP \\
    29 & 71 & 2 & LP \\
    30 & 86, 135, 149 & 3 & C \\
    32 & 251 & 1 & 1 \\
    33 & 123 & 2 & LP \\
    34 & 48, 187, 243 & 2 & LP \\
    35 & 49, 59, 115 & 2 & LP \\
    36 & 213 & 2 & LP \\
    37 & 91 & 2 & LP \\
    38 & 52, 155, 211 & 2 & LP \\
    40 & 96, 235, 249 & 1 & 1 \\
    41 & 97, 107, 121 & 4 & T \\
    42 & 112, 171, 241 & 2 & LP \\
    43 & 113 & 2 & LP \\
    44 & 100, 203, 217 & 2 & LP \\
    45 & 75, 89, 101 & 3 & C \\
    46 & 116, 139, 209 & 2 & LP \\
    50 & 179 & 2 & LP \\
    51 & - & 2 & LP \\
    54 & 147 & 4 & T \\

    \end{tabular}
    \begin{tabular}{|c|c|c|c}

        Rule & Equivalent & Class & Subclass \\
        \hline
        56 & 98, 185, 227 & 2 & LP \\
        57 & 99 & 2 & LP \\
        58 & 114, 163, 177 & 2 & LP \\
        60 & 102, 153, 195 & 3 & U \\
        62 & 118, 131, 145 & 2 & HP \\
        72 & 237 & 2 & LP \\
        73 & 109 & 2 & T \\
        74 & 88, 173, 229 & 2 & LP \\
        76 & 205 & 2 & LP \\
        77 & - & 2 & LP \\
        78 & 92, 141, 197 & 2 & LP \\
        90 & 165 & 3 & U \\
        94 & 133 & 2 & LP \\
        104 & 233 & 2 & LP \\
        105 & - & 3 & U \\
        106 & 120, 169, 225 & 4 & C \\
        108 & 201 & 2 & LP \\
        110 & 124, 137, 193 & 4 & T \\
        122 & 161 & 3 & C \\
        126 & 129 & 3 & C \\
        128 & 254 & 1 & 1 \\
        130 & 144, 190, 246 & 2 & LP \\
            132 & 222 & 2 & LP \\
        134 & 148, 158, 214 & 2 & LP \\
        136 & 192, 238, 252 & 1 & 1 \\
        138 & 174, 208, 224 & 2 & LP \\
        140 & 196, 206, 220 & 2 & LP \\
        142 & 212 & 2 & LP \\
        146 & 182 & 3 & C \\
        150 & - & 3 & U \\
        152 & 188, 194, 230 & 2 & LP \\
        154 & 166, 180, 210 & 2 & LP \\
        156 & 198 & 2 & LP \\
        160 & 250 & 1 & 1 \\
        162 & 176, 186, 242 & 2 & LP \\
        164 & 218 & 2 & LP \\
        168 & 224, 234, 248 & 1 & 1 \\
        170 & 240 & 2 & LP \\
        172 & 202, 216, 228 & 2 & LP \\
        178 & - & 2 & LP \\
        184 & 226 & 2 & LP \\
        200 & 236 & 2 & LP \\
        204 & - & 2 & LP \\
        232 & - & 2 & LP \\
    \end{tabular}
\end{table*}

\end{document}